# Practical free-space quantum key distribution over 10 km in daylight and at night

Richard J. Hughes, Jane E. Nordholt, Derek Derkacs and Charles G. Peterson

*Physics Division, Los Alamos National Laboratory, Los Alamos, NM 87545, USA*

We have demonstrated quantum key distribution (QKD) [1] over a 10-km, 1-airmass atmospheric range during daylight and at night. Secret random bit sequences of the quality required for the cryptographic keys used to initialize secure communications devices were transferred at practical rates with realistic security. By identifying the physical parameters that determine the system's secrecy efficiency, we infer that free-space QKD will be practical over much longer ranges under these and other atmospheric and instrumental conditions.

Cryptography allows two parties ("Alice" and "Bob") to render their communications unintelligible to a third party ("Eve"), provided they both possess a secret random bit sequence, known as a cryptographic key, which is required as an initial parameter in their encryption devices [2]. Secure key distribution is then essential; Eve must not be able to obtain even partial knowledge of the key. Key distribution using a secure channel ("trusted couriers") is effective but cumbersome in practice, potentially vulnerable to insider betrayal and may not even be feasible in some applications. In contrast quantum key distribution (QKD) [1] uses single-photon communications to generate and transfer new keys on-demand with security based on fundamental quantum principles in concert with information-theoretically secure protocols [3]; Eve can do no better than to guess the key. QKD offers long-term security superior to public-key based key transfer systems [4]. QKD would be especially useful if it could be performed reliably across line-of-sight paths through the atmosphere [4, 5]. Free-space QKD has previously been demonstrated over laboratory [6, 7] and modest outdoor [8, 9, 10] distances. More recently, the *feasibility* of free-space QKD over kilometer-scale distances has been demonstrated in both daylight [11] and at night [11, 12]. In this paper we report the first demonstration of the transfer of cryptographic quality secret keys at practical rates during both day and night using QKD across a 10-km air path whose extinction, optics and background are representative of potential applications. We also develop a methodology for extrapolating these results to other ranges under other atmospheric and instrumental conditions.

In our realization of the "BB84" QKD protocol [1] Alice (the transmitter) sends a sequence of random bits over a "quantum channel" to Bob (the receiver) that are randomly encoded as linearly polarized single photons in either of two conjugate polarization bases with $(0, 1) = (H, V)$, where "H" ("V") denotes horizontal (vertical) polarization (respectively), in the "rectilinear" basis, or $(0, 1) = (+45º, -45º)$, where "+45º" and "-45º" denote the polarization directions in the "diagonal" basis. Bob randomly analyzes the polarization of each arriving photon in either the (H, V) or the (+45º, -45º) basis, assigning the corresponding bit value to detected photons. Then using a "public channel", which is authenticated but assumed to be susceptible to passive monitoring by Eve, he informs Alice in which time slots he detected photons, but without revealing the bit value he assigned to each one. The sequence of bits detected by Bob and the corresponding bit sequence transmitted by Alice form partially correlated "raw" keys. (See Figure 1.) Then using the "public channel" Alice reveals her basis choice for each bit of her raw key, but not the bit value. Bob communicates back the time slots of the bits in his raw key for which he used the same basis as Alice. In an ideal system Alice's transmitted bits and the results of Bob's measurements on this random, 50% portion of the raw key, known as the "sifted" key, are perfectly correlated; they discard the raw key bits for which Bob used the wrong basis. In practice Bob's sifted key contains errors. Fundamental quantum principles ensure that Eve is both limited in how much information she may obtain by eavesdropping on the quantum communications, and that she cannot do so without introducing a disturbance (errors) in Bob's sifted key from which Alice and Bob can deduce a rigorous upper bound on leaked information. Alice and Bob determine this bound after reconciling their sifted keys using *post facto* error correction [13] over their public channel, but at the price of leaking additional (side) information to Eve. From their partially-secret reconciled keys Alice and Bob extract the shorter, final secret key on which they agree with overwhelming probability and on which Eve's expected information is much less than one bit [14] after a final stage of "privacy amplification" [3] using further public channel communications. BB84 with ideal single-photon signals is unconditionally secure [15].

The atmosphere is not birefringent at optical wavelengths and so can function as a quantum channel for the transmission of BB84 polarized single-photon states. Atmospheric transmittance and the availability of high-efficiency, low-noise single-photon detectors (SPDs) strongly constrain the operational wavelength, with 772-nm offering the highest secret bit rates with current technology [5]. Challenges to implementing free-space QKD include the background radiance, which is a strong error source even at night [11], that varies over several orders of magnitude on time scales of the order of hours, and atmospheric turbulence which introduces random variations in the quantum channel transmission on 10-100-ms time-scales. These features of the free-space quantum channel also present challenges to extrapolating the performance of QKD from results at one range and one time-of-day to other ranges and other times-of-day. Our implementation of free-space QKD effectively deals with the physics challenges of the atmospheric quantum channel and we have developed a formalism that allows system performance to be extrapolated into other regimes.

Our free-space QKD system uses spectral, spatial and temporal filtering to render the background tractable [4, 5, 10, 11]. It has no active polarization switching elements, both as a security feature and for design simplicity, and can operate across ranges up to 30 km (limited by the range of our 1 Mbit s$^{-1}$ wireless Ethernet "public channel"). On each cycle of a 1-MHz clock the transmitter (Alice) emits a $\sim$ 1-ns, few mW, 1,550-nm timing pulse. After a 100-ns delay, two secret random bits generated by a cryptographic monolithic randomizer [16] determine which one of four temperature-controlled "data" diode lasers emits a $\sim$ 1-ns, 772-nm optical pulse with one of the BB84 polarizations [1] and an average photon number, $\mu$, (we assume Poissonian photon number statistics and $\mu < 1$ throughout) that is launched towards the receiver (Bob). (See Figure 2.) At Bob the timing pulse is detected by a photodiode, to set up an $\sim$ 1-ns timing "slot" in which a QKD data pulse is expected. An 18-cm Cassegrain telescope, whose field of view is restricted to $\sim$ 220-$\mu$rad by a spatial filter, collects the 772-nm data pulse, passes it through a 0.1-nm wide interference filter (transmission $\eta_{filt} \sim 0.6$), and directs it into an optical system where its polarization is randomly analyzed in one of the BB84 bases. SPDs, one for each of the four BB84 polarizations, register them. (The SPDs are based on passively-quenched EG&G model #C30902S silicon avalanche photodiodes operated at a temperature of -20°C, with a single-photon detection efficiency of $\eta_{det} \sim 0.61$, and a dark count rate of $\sim$ 1.6 kHz. Lower dark count rates $\sim$ 100 Hz would be possible with other detectors.) After a 1-s quantum transmission of $10^6$ bits, a 6-s public channel communication is required to produce the sifted key. Cameras in Alice and Bob provide rudimentary visual authentication to protect against a "man-in-the-middle" attack. Cryptographic authentication [2, 14, 17] would be included in a complete QKD-enabled secure communications system.

We located Alice at an elevation of 2,760 m on Pajarito Mountain, Los Alamos, NM, (35° 53.489' N, 106° 22.647' W) with Bob located near to our laboratory (35° 52.222' N, 106° 16.312' W), at an elevation of 2,153 m, pointing towards Alice (azimuthal direction = 284° true, elevation angle = 3.5°). The 9.81-km Alice-Bob air path had an average beam height above the terrain of $\sim$ 140 m and a calculated atmospheric transmittance of $\eta_{trans} = 0.81$ [18]. We operated the system for several hours on each of several days during both full daylight, with $0.2 < <\mu>_{day} < 0.8$, and at night with $0.1 << \mu>_{night} < 0.2$. During a 1-s quantum transmission, the probability for a transmitted bit to enter the sifted key, $P_{sif}$, depends on the average photon number of the optical pulses, $\mu$, the efficiency of the atmospheric transmission and the receiver's detection efficiency. In the regime in which we operate, where the signal-to-background ratio is large and the probability of multi-detection events is much less than the probability of single-detection events, we may write, $P_{sif} \approx \left[ 1 - \exp\left( -\mu \eta_{trans} \eta_{geo} \eta_{rec} \eta_{filt} \eta_{BB84} \eta_{det} \right) \right]$, where $\eta_{geo}$ ($\sim 3 - 12\%$) is the 1-s average geometric capture efficiency of data pulses by Bob, and $\eta_{rec} \sim 0.47$ is the transmission of Bob's receiver optics with the exception of the 50/50 beamsplitter that provides the random choice of QKD polarization measurement basis, whose transmission/reflection coefficient is $\eta_{BB84} = 0.5$. Taking into account the 1-MHz clock rate, the system produced $n = 10^6 P_{sif} \sim 100 - 2,000$ sifted key bits per 1-s quantum transmission. Errors in Bob's sifted key were overwhelmingly caused by (unpolarized) background photons in daylight and by detector dark noise at night. We quantified these sources of errors by operating the system at $\mu = 0$ (zero transmitted photon number) to produce a "sifted key" formed entirely from detections of background photons and detector dark counts at Bob's receiver. We found that in each 1-s, $\mu = 0$ transmission, each of Bob's four detectors registered approximately equal numbers of detections, of which approximately one half were in the "wrong" basis, and the balance of the detections contributed bits to a sifted key that were divided roughly equally between "correct bits" and "errors". With both transmitter and receiver in afternoon sunlight we observed $C \sim 50$ sifted key errors per detector in a 1-s, $\mu = 0$ transmission, corresponding to a radiance of $\sim 2$ mW cm$^{-2}$ $\mu$m$^{-1}$ str$^{-1}$. (Under these conditions Alice and Bob produced background-generated sifted keys containing $\sim 400$ bits of which $\sim 200$ of the bits in Bob's sifted key were errors.) In "reduced daylight" (transmitter in shadow, receiver in direct sunlight) this dropped to $C \sim 5$. At night, even though the background radiance is at least a factor of one million less than in daylight, we found $C \sim 1 - 2$, owing to detector dark noise. Therefore, for $\mu \neq 0$ transmissions, the sifted key bit error rate (BER) can be written as $\varepsilon \approx 4C/n \approx CD/\mu \eta_{opt}$, where $\eta_{opt} = \eta_{trans} \eta_{geo}$ is the atmospheric quantum channel's 1-s average efficiency, $C$ has the value appropriate to the time-of-day of the transmission as described above, and $D \approx 4 \times 10^{-6}/\eta_{rec} \eta_{filt} \eta_{BB84} \eta_{det} \approx 4.7 \times 10^{-5}$ is a constant, receiver-dependent factor. (Errors caused by polarization misalignments and imperfections were estimated to contribute $< 0.5\%$ to the sifted key BER.) We note that the sifted key BER, which is one of the relevant quantities in determining the overall QKD system performance, is a function of: the average photon number, $\mu$, which is characteristic of the transmitter; the quantity $\eta_{opt}/C$, which characterizes the quality of the atmospheric quantum channel; and the constant, receiver-dependent quantity, $D$. In what follows we will find that $\mu$ and $\eta_{opt}/C$ are particularly useful independent variables for predicting system performance. (The quantity $\eta_{opt}/C$ is proportional to the signal-to-noise ratio that the system would have for producing sifted bits, scaled to a notional photon number of $\mu = 1$. We have chosen to use $\eta_{opt}/C$ as an independent variable because this quantity isolates the dependence of the system's performance on the atmospheric channel properties, which are not under our direct control, into a single quantity whose value is determined from measured quantities.)

For example, on 4 October 2001 under cloudless New Mexico skies between 17:42 MDT, with Bob in direct sunlight, and sunset at 18:44 MDT we made 207 1-s quantum transmissions (207 million random bits transmitted by Alice) that had an average photon number of $<\mu>_{day} \approx 0.49$ with a standard deviation of 0.12, a channel efficiency of $<\eta_{opt}>_{day} = 4.1\% \pm 1.2\%$, and channel parameter $<\eta_{opt}/C>_{day} = 0.0026 \pm 0.0017$. From these we obtained 394,004 sifted key bits ($<P_{sif}>_{day} = (1.9 \pm 0.8) \times 10^{-3}$), with an average BER of $<\varepsilon>_{day} = 5.0\% \pm 2.2\%$. (See Table 1 for an example of a daylight sifted key.) Then between 18:44 MDT and 19:29 MDT from a further 236 1-s quantum transmissions with an average photon number of $<\mu>_{night} \approx 0.14$ with a standard deviation of 0.02, a channel efficiency of $<\eta_{opt}>_{night} = 6.6\% \pm 1.8\%$, and channel parameter $<\eta_{opt}/C>_{night} = 0.017 \pm 0.007$, we obtained 192,925 sifted key bits ($<P_{sif}>_{night} = (0.82 \pm 0.21) \times 10^{-3}$), with an average BER of $<\varepsilon>_{night} = 2.1\% \pm 0.7\%$. (See Figure 3.)

Alice and Bob reconcile their $n$-bit sifted keys from each 1-s quantum transmission using the interactive "bisective search" algorithm [6, 13] to correct Bob's errors by dividing the sifted key from each 1-s quantum transmission into words and the parity of each word is publicly communicated. Words whose parities do not match are then repeatedly sub-divided and the parity of the subwords publicly communicated to locate and correct an error. The key is then randomly shuffled and the process repeated until no parity mismatches occur on two successive rounds. Alice and Bob then possess *n*-bit reconciled keys that agree with very high probability; they have a reliable estimate of the BER of the sifted key, but they have revealed parity ("side") information about the sifted key that is approximately 19% greater than the Shannon limit of $f(\varepsilon) = -\varepsilon \log_2 \varepsilon + (1-\varepsilon) \log_2 (1-\varepsilon)$ bits per bit of sifted key, on sifted keys of $\sim 10^4$ bits for the BERs we encounter.

Our first line of defense for Alice and Bob against eavesdropping is similar to that of Reference [6]. To protect against opportunities presented by multi-photon signals [19] (e.g. a beamsplitting attack) Alice and Bob assume that the fraction $(\approx \mu)$ of sifted key bits in each 1-s transmission that originated from the transmitter as multi-photon pulses could have been faithfully identified by Eve [6]. They attribute all Bob's errors to Eve having performed an intercept/resend attack in the Breidbart basis on the portion $\approx (4\varepsilon/1-\mu)$ of sifted key bits that originated as single-photon pulses [6]. The number of secret bits that Alice and Bob can extract from *n* reconciled key bits is then $F(n,\mu,\varepsilon) \approx \lfloor n[R(\mu,\varepsilon) - 1.19 f(\varepsilon)] - s \rfloor$ bits [20], where *s* is a safety factor [3, 14], and $R(\mu,\varepsilon) \approx 1 - \mu - 4\varepsilon \log_2 (1.5)$ is Eve's collision entropy per bit [3] of the sifted key. (If the reconciled key has any "bias" − more "zeroes" than "ones" or *vice-versa* − they also reduce the collision entropy appropriately to compensate for the information that this would provide to Eve.) From each positive-*F* 1-s quantum transmission Alice and Bob produce an *F*-bit final secret key using privacy amplification by public communications [3]. They form the elements of their final secret keys as the parities of *F* random (but publicly specified) subsets of their *n*-bit reconciled keys.[i] The protocol fails to produce a secret key if *F* < 0. Eve's expected (Shannon) information on the final key $(< 2^{-s}/\ln 2 \text{ bits})$ is < $10^{-6}$ bits for *s* = 20, independent of the length of the final key. We define two figures of merit: the "privacy amplification efficiency" (the number of secret bits per sifted bit), $P_{sif \rightarrow secret} = F/n$, for *F* > 0, and 0 otherwise; which characterizes the efficiency of the information-theoretic parts of the QKD procedure, and the "secrecy efficiency" (the number of final secret bits per transmitted bit), $P_{secret} = P_{sif} \, P_{sif \rightarrow secret}$ for *F* > 0, and 0 otherwise, which characterizes the performance of the entire QKD system. (See Figure 1.) In our system, which operates at a 1-MHz clock rate, the total number of secret bits that can be produced from a 1-s quantum transmission is therefore $10^6 P_{secret}$.

First we consider $P_{sif \rightarrow secret}$, which depends only on $\mu$ and $\eta_{opt}/C$ in the regime in which we operate where the safety factor, *s*, is much less than the number of sifted key bits, *n*. Remarkably, only certain ranges of $\mu$ and $\eta_{opt}/C$ values allow *any* secret bits to be extracted from the corresponding sifted keys [11]: no secret bit yield is possible for channel parameters smaller than $\eta_{opt}/C = 0.0016$ for any value of $\mu$ with our system. (See Figure 4.) Of the 207 daylight 1-s quantum transmission on 4 October, 2001, 94 lie in this zero-yield region. For larger values of $\eta_{opt}/C$ there is a range of $\mu$-values, $\mu_{min} < \mu < \mu_{max}$, (where $\mu_{min}$ and $\mu_{max}$ are functions of $\eta_{opt}/C$) consistent with non-zero secret bit yield. For $\mu < \mu_{min}$ the sifted key BER is so large that no secret bits can be extracted because of the large amount of information potentially leaked to correct errors and through intercept/resend eavesdropping. As $\mu$ increases from $\mu_{min}$ the sifted key BER decreases, and the yield of secret bits initially increases, but as $\mu$ increases further so much information is potentially available to Eve through multi-photon pulses that the secret bit yield starts to decrease, reaching zero at some value $\mu_{max}$. The remaining 113 of our daylight 1-s transmissions and all of our night transmissions lie in this allowed region, which shrinks to zero for $\eta_{opt}/C = (\eta_{opt}/C)_{min} = 0.0016$ at $\mu \approx 0.45$ with $\varepsilon \approx 5.7\%$. This observation allows us to specify the limiting atmospheric channel conditions under which QKD is possible with this system (and for other systems by scaling the relevant parameters) and as we will see later, to infer the maximum range under different background conditions. The average privacy amplification efficiencies for extracting secret bits from the 4 October sifted keys were: $<P_{sif \rightarrow secret}>_{day} = 0.26 \pm 0.12$ for the non-zero secret bit yield daylight transmissions only; $<P_{sif \rightarrow secret}>_{day} = 0.06 \pm 0.25$ when the zero-yield daylight transmissions were included; and $<P_{sif \rightarrow secret}>_{night} = 0.64 \pm 0.07$ for the night transmissions.

---

[i] With very small probability, two errors may remain in Bob's reconciled key after error correction. Privacy amplification then has the effect of producing final keys in which half of Bob's bits disagree with Alice's. Obviously, such keys cannot be used. Fortunately, this rare occurrence can be detected with high probability by performing a final key check, in which we sacrifice a few final key bits and compare them to ensure that the keys agree, leading to a small reduction in the overall secrecy efficiency. This key check would be included in the authentication procedure [17] in a complete system [14], but this was not implemented when the data in this paper was taken.

The secrecy efficiency $P_{secret}$, is the most relevant figure of merit for overall system performance. This quantity determines the total number of secret bits that Alice and Bob can produce per unit time, and through its dependence on the relevant independent parameters we can determine how to optimize its value. In daylight we achieved a maximum secrecy efficiency of $P_{secret, max} = 7.0 \times 10^{-4}$, an average value of $<P_{secret}>_{day} = (3.2 \pm 1.4) \times 10^{-4}$ for the non-zero yield transmissions, and $<P_{secret}>_{day} = 1.5 \times 10^{-4}$ including all daylight transmissions. The corresponding values at night were $P_{secret, max} = 8.0 \times 10^{-4}$ and $<P_{secret}>_{night} = (4.2 \pm 1.4) \times 10^{-4}$. (See Figure 5 and Table 1 for an example of a final secret key.) The total 50,783 of daylight final secret key bits, and the total 118,064 of night final secret key bits passed the FIPS 140-2 cryptographic randomness tests [21] as well as the 5-bit version of the Maurer universal statistical test for cryptographic random numbers [22]. The FIPS tests, which we also apply to the random numbers produced by Alice's randomizer, require samples containing 20,000 bits and specify statistical significance levels for: the proportions of 1's ("monobit test"); the frequencies with which all possible four-bit groups occur ("poker test"); and the frequencies with which consecutive sequences of 0's ("gaps") and 1's ("runs") occur ("runs test"). Maurer's test, which requires large samples of bits, sets statistical significance levels for the intervals between repetitions of $m$-bit blocks of bits. Our 4 October 2001 data provided enough secret bits to perform this test for $m = 5$.

We also consider protecting Alice and Bob from two eavesdropping attacks in which Eve would take complete control of the atmospheric quantum channel. First, we consider the possibility that Eve could perform a technologically-feasible version of an unambiguous state discrimination (USD) attack [6, 23, 24] to uniquely identify the polarization of a portion of the optical pulses containing three photons emerging from Alice's transmitter. Eve could couple all of Alice's optical pulses with perfect optical efficiency into a lossless version of Bob's receiver. Whenever precisely three of Eve's single-photon detectors are triggered, she can uniquely identify the pulse's polarization as the polarization associated with the single detector that was triggered in one of the bases. Using a conventional channel Eve could then communicate the polarization to a transmitter similar to Alice's located adjacent to Bob's receiver, and fabricate an optical pulse of the same polarization. Eve would simply block all other data pulses. Eve would then know precisely every bit in Alice's sifted key and would be able to evade detection provided she did not reduce Bob's raw key rate below the expected value. This is only possible if the emission rate of three-photon pulses from the transmitter that Eve can identify is larger than the expected single-photon arrival rate at the receiver: $\mu^2 / 32 > \eta_{opt}$. None of our positive secrecy efficiency data lies in this region and so our data is secure against this attack. (Our system can tolerate up to 21dB of atmospheric channel loss at $\mu = 0.5$, and up to 31dB of loss for $\mu = 0.15$ while maintaining security against this form of eavesdropping.) Second, we consider the possibility of an even stronger, photon number splitting (PNS) attack. In the version of the attack that we consider Eve would block all single photon pulses from Alice, split off and store one photon from each multi-photon pulse while sending on the remaining photons to Bob over a lower-loss channel, and then measure the polarization of her stored photon once Alice announces her basis choices. (We do not consider the possibility that Eve could *increase* the efficiency of Bob's detectors [25].) As with the USD attack considered above, Eve would then know precisely every bit in Alice's sifted key and would be able to evade detection provided she did not reduce Bob's raw key rate below the expected value. This would be feasible if the emission rate of multi-photon pulses from the transmitter is larger than the single-photon arrival rate at the receiver ($\mu > 2\eta_{opt}$) [25], which imposes a stronger limit on the allowable photon numbers than for the USD attack. However, Eve would require yet-to-be-invented technology: an optical-photon-number quantum non-demolition measurement capability, a quantum memory and a lower-loss quantum channel to Bob. Nevertheless, approximately half of the 4 October, 2001 night 1-s transmissions (whose 107,250 sifted bits yielded 70,577 final secret bits) are secure against this version of the PNS.

We have demonstrated that free-space QKD is possible in daylight or at night, protected against intercept/resend, beamsplitting and USD eavesdropping (and even PNS eavesdropping at night), over a 10-km, 1-airmass path, which is representative of potential ground-to-ground applications and is several times longer than any previously reported results. Our system provided cryptographic quality secret key transfer with a number of secret bits per 1-s quantum transmission that would support practical cryptosystems such as the Advanced Encryption Standard, AES [26], or "one-time pad" encryption for short messages. (See Figure 6 for an example.) We have also developed a methodology that allows us to deduce the secrecy efficiency for other transmission distances, instrumental conditions, atmospheric properties and radiances by scaling the $\eta_{opt}/C$ parameter from its 10-km values, and by noting that the quantity $P_{secret}/\eta_{opt}$ is a function of $\mu$ and $\eta_{opt}/C$ only. (See Figure 7.) First, since no secret bits can be produced for $\eta_{opt}/C < (\eta_{opt}/C)_{min} = 0.0016$, we infer that free-space QKD would be feasible *with this system*, at reduced rates, over high-desert ground-to-ground atmospheric paths of up to 15 km in full daylight, 30 km in "reduced daylight" (transmitter in shadow) and 45 km at night. Second, optimal secrecy efficiency is attained for $\mu \sim 0.5$, independent of range and time-of-day when the USD and PNS eavesdropping possibilities are not considered. Third, our methodology allows us to infer the performance gains that could be expected from various instrumental changes. Implementation of fast pointing beam control at the transmitter is likely to increase the value of $\eta_{opt}$, and hence $\eta_{opt}/C$ which would increase the number of secret bits that could be created per unit time at a given range and the system's maximum range. In contrast, although increasing the receiver aperture would also increase the value of $\eta_{opt}$, and hence the secret bit rate at a given range, it would *not* increase the maximum *daylight* range because the value of $\eta_{opt}/C$ would be unchanged. This is because in daylight the quantity $C$ is background dominated, and the increased receiver aperture would increase this quantity in proportion to the increase in $\eta_{opt}$. However, an increased receiver aperture would increase the maximum range of our system at *night*, because $C$ is then dominated by detector dark noise, which would not be altered by the aperture increase. The use of lower-noise SPDs would also allow higher secret bit yields and longer ranges at night because of the reduction in $C$ and increase in $\eta_{opt}/C$. An improvement in the error correction efficiency would allow modest improvements in both the secret bit yield and range. Finally we believe that the methodology that we have developed for relating the overall system performance to instrumental and quantum channel properties may also be applicable to other QKD systems, including optical-fiber based ones.

Peter Dickson and the board of the Los Alamos Ski Club and the National Forest Service are thanked for providing access to their property for this experiment. National Reconnaissance Office Director's Innovation Initiative funding administered by Col. John Comtois and Peter Hendrickson is gratefully acknowledged. It is a pleasure to thank George Morgan and Christopher Wipf for helpful discussions.

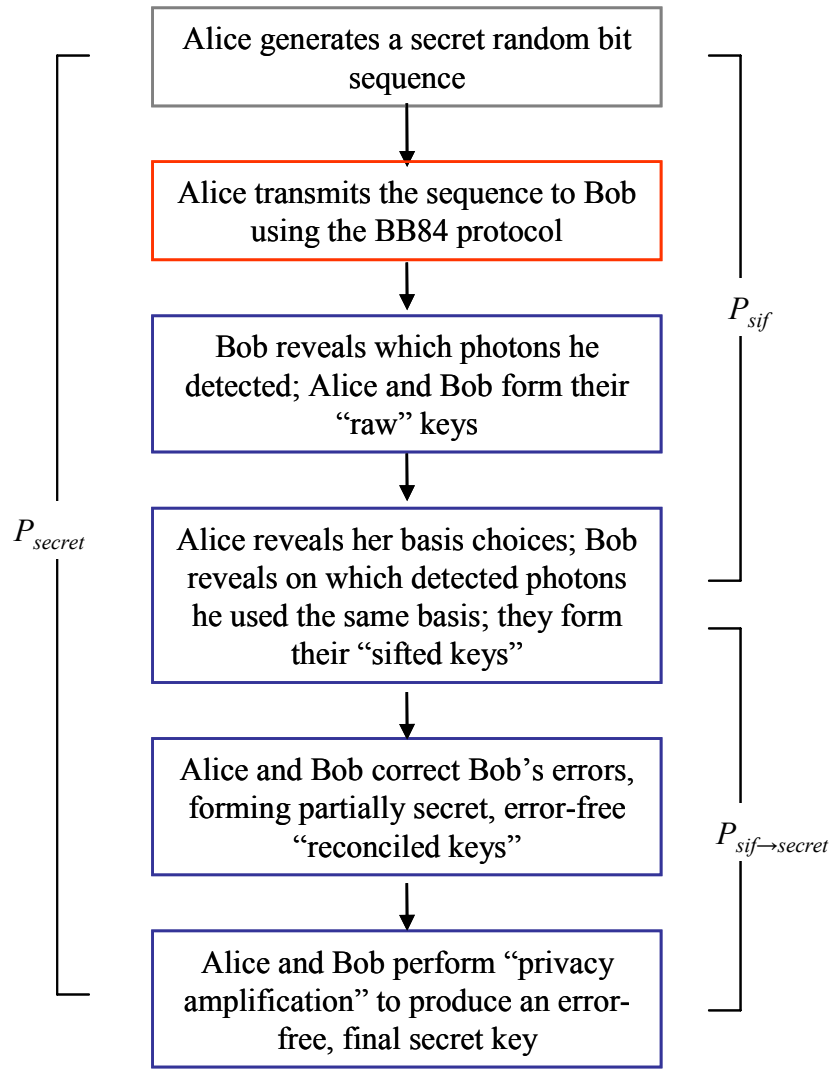

Figure 1. This figure shows the sequence of events in a QKD procedure leading to a cryptographic key shared by Alice and Bob on which they agree with overwhelming probability and on which Eve knows very much less than one bit of information. Also shown are three figures-of-merit for characterizing the process: $P_{sif}$, which characterizes the efficiency with which sifted bits are produced from the initial bit sequence; $P_{sif \rightarrow secret}$, which characterizes the efficiency of extracting secret bits from the sifted bits; and $P_{secret}$, which characterizes the overall efficiency for generating final secret bits. (See text for details.)

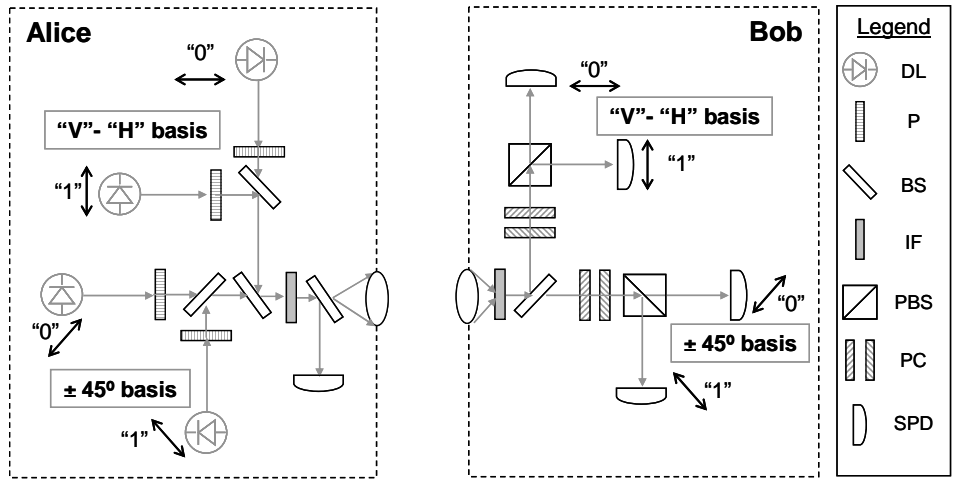

Figure 2. The polarization optics of the QKD transmitter and receiver. The outputs of the data lasers (DL) in Alice are attenuated (average photon number $\mu < 1$), their polarizations set to the BB84 values (shown as two-headed arrows) by linear polarizers (P), combined using beamsplitters (BS), passed through a spatial filter to erase spatial mode information (not shown), constrained by an interference filter (IF) to remove spectral information and then directed onto a BS. Photons transmitted through the BS are launched towards Bob, whereas those reflected are directed onto a single-photon detector (SPD) with a $\sim 20$ ns timing window, to monitor the $\mu$-value of the launched data pulses. The relative timings of the DLs are matched to within the SPD timing jitter. At Bob data pulses pass through an IF and onto a BS where they are randomly transmitted or reflected. Along the reflected path, a data pulse's polarization is analyzed in the rectilinear basis, using a polarization controller (PC) and a polarizing beamsplitter (PBS). If one of the SPDs in the PBS output ports fires within the timing window (and no other SPD fires) Bob assigns a bit value to the data pulse. An analogous procedure occurs for data pulses taking the transmitted path where they are polarization analyzed according to Bob's conjugate (diagonal) basis. (We estimate that the probability for a photon produced in the SPD breakdown "flash" [27] to emerge from the receiver telescope is $< 10^{-9}$.) Multi-detection events, in which more than one SPD fires, are recorded but not used for key generation.

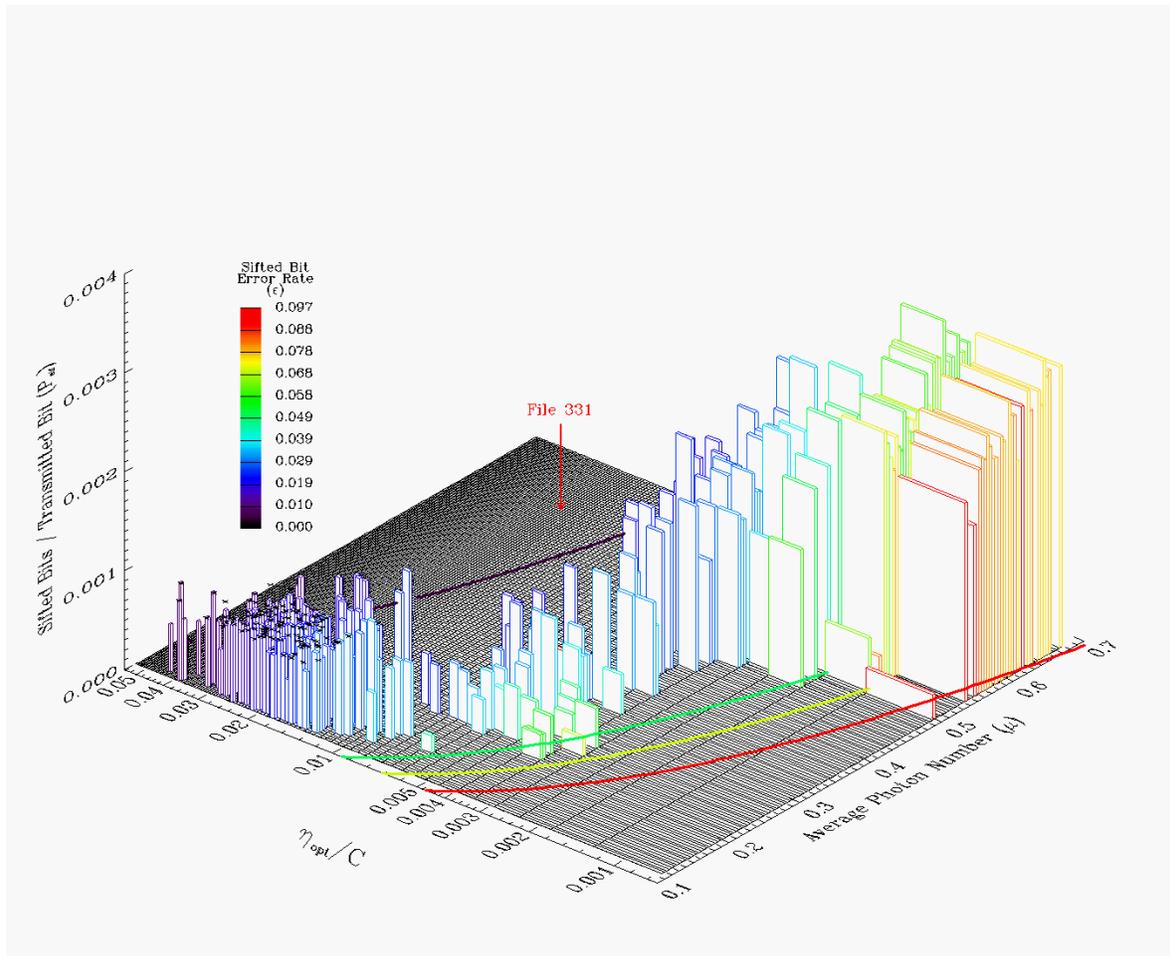

Figure 3. A histogram of the efficiency with which sifted key bits were transferred ($P_{sif}$) in 1-s quantum transmissions on 4 October, 2001 versus the average photon number μ and the atmospheric quantum channel parameter, $\eta_{opt}/C$, values and color-coded by the sifted key BER, ε. Several 1-s, 10-km transmissions are grouped into each vertical column. The solid lines are contours of constant sifted key BER, with their color indicating the corresponding ε value: the purple contour has ε = 0.5%; the green contour has ε = 5%; the yellow contour has ε = 7%; and the red contour has ε = 10%. The transmissions with $\eta_{opt}/C < 0.01$ were made in daylight. The transmission marked by the red arrow ("File 331") is described in detail in Table 1. The transmissions marked with asterisks are PNS-safe night transmissions. (See text for details).

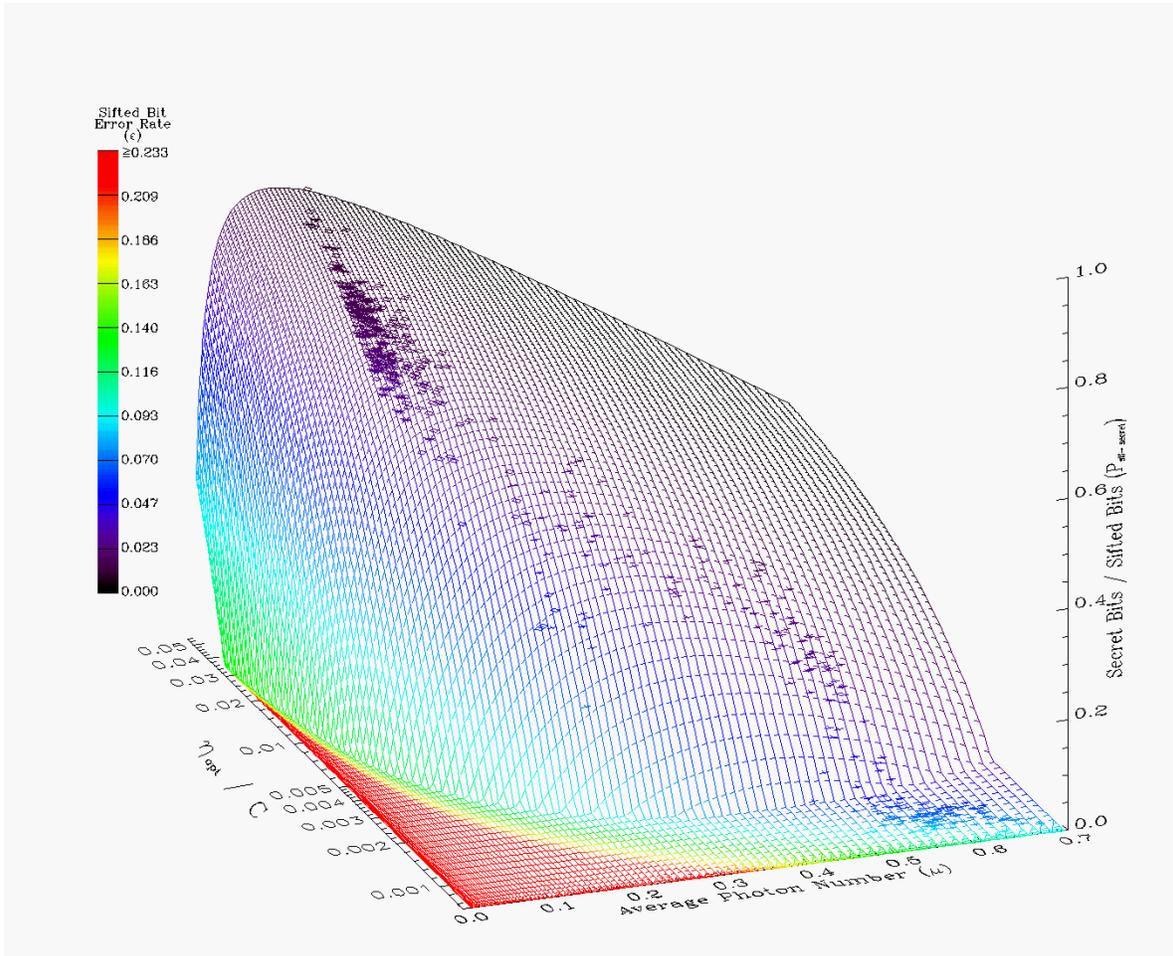

Figure 4. A surface plot showing the privacy amplification efficiency $P_{sif \rightarrow secret}$, with which secret bits can be extracted from a sifted key for our system, which is a function of two independent variables: the average photon number, μ, and the atmospheric quantum channel parameter, $\eta_{opt}/C$. The locations of 1-s, 10-km quantum transmissions from 4 October, 2001 are marked on this surface, which is color coded by the sifted key BER. The privacy amplification efficiency for other ranges and conditions is given by the point on this surface, whose location is specified by the $\eta_{opt}/C$ value, which may be obtained by scaling from the 10-km values, and the average photon number, μ. Where $P_{sif \rightarrow secret}$ drops to zero, no secret bits can be extracted from the sifted key. (See text for details.)

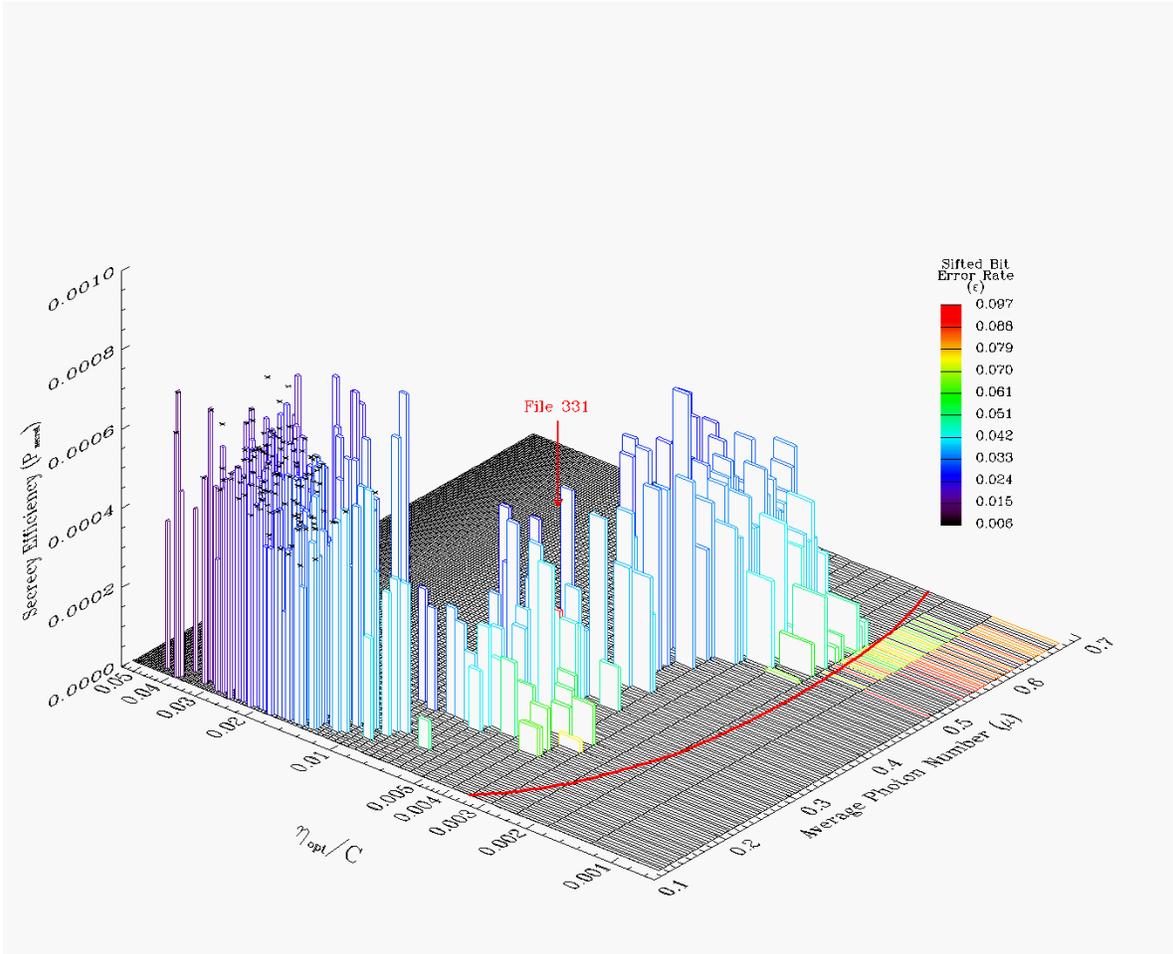

Figure 5. A histogram of the secrecy efficiency, $P_{secret}$, (the number of secret bits produced per transmitted bit) versus the average photon number, $\mu$, and the atmospheric quantum channel parameter, $\eta_{opt}/C$, values color-coded by the sifted key BER, $\varepsilon$, of 1-s, 10-km quantum key transmissions on 4 October, 2001. Several 1-s transmissions are grouped into each vertical column. The key transmission marked with the red arrow is described in Table 1. In the region below the red line no secret bits can be transferred with this system. For example, we see that a portion of our daylight data (with $\eta_{opt}/C < 0.0016$) lies in this region. Even though these transmissions yielded a large number of sifted bits (see Figure 3), no secret bits could be produced from them after reconciliation and privacy amplification (see Figure 4) because of the large amount of information about the reconciled key that Eve could have acquired from eavesdropping on the quantum transmissions and passively monitoring the public channel error correction transmissions between Alice and Bob. The night transmissions marked with asterisks are PNS-safe. (See text for details).

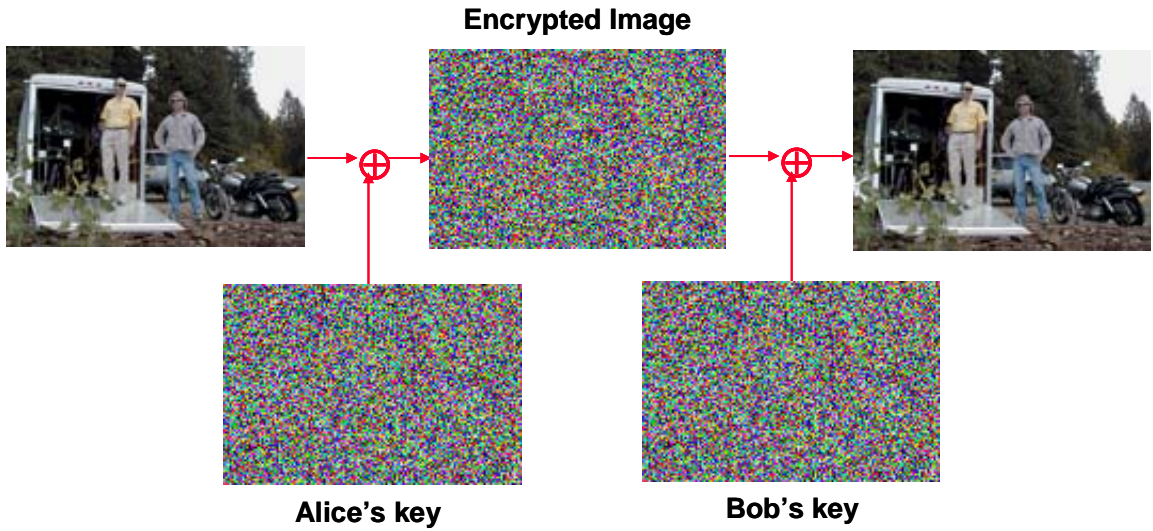

**Encrypted Image**

**Alice's key**

**Bob's key**

Figure 6. An example of secure communications using a one-time pad constructed from cryptographic key material produced by 10-km free-space QKD. The digital image at left, which shows two of the authors (RJH and CGP) standing next to the free-space QKD transmitter (Alice) at one end of the 10-km range, is composed of 140x94 12-bit color pixels. Each bit of the image was encrypted by XOR-ing it with a secret key bit produced by QKD to produce an encrypted image, which was then communicated to Bob over a public channel, requiring 157,920 key bits in total. Eve would not be able to discern the original image through the randomization introduced by the encryption, but Bob can recover the image by XOR-ing each bit of the encrypted image with the appropriate bit of his secret key. Alice's and Bob's keys are represented as random images in which each pixel is the RGB representation of 12 bits of key. This one-time pad encryption is unconditionally secure but requires as many secret key bits as message bits. Practical cryptosystems would only need a few hundred secret key bits to encrypt large quantities of data.

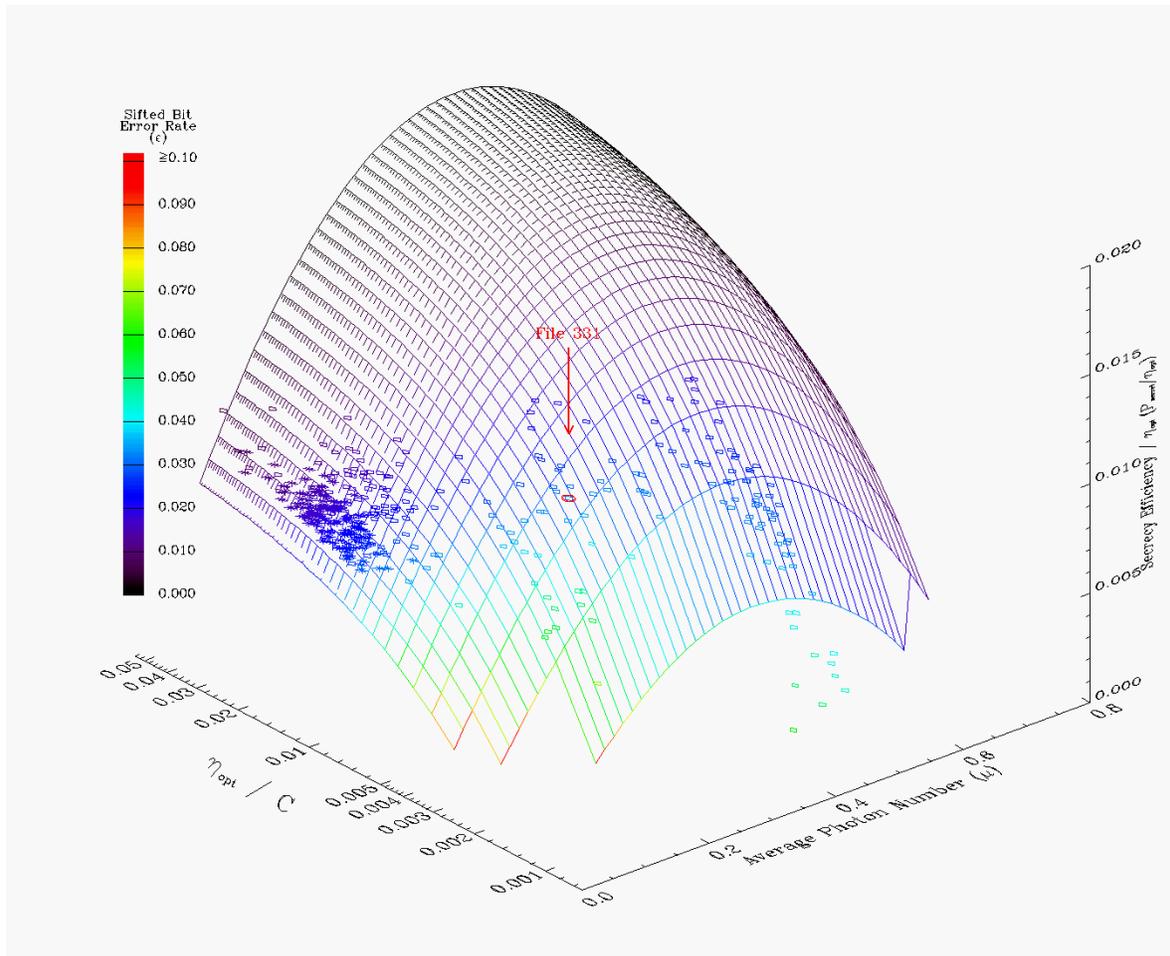

Figure 7. A surface plot of the secrecy efficiency, $P_{secret}$, scaled by the atmospheric quantum channel's efficiency, $\eta_{opt}$, versus the average photon number, $\mu$, and the quantum channel parameter, $\eta_{opt}/C$, for our system. The locations of 1-s, 10-km quantum transmissions from 4 October, 2001 are marked on this surface, which is color coded by the sifted key BER. The data indicated by the red arrow ("File 331") is described in detail in Table 1. The secrecy efficiency for other ranges and conditions is given by the point on this surface, whose location is specified by the $\eta_{opt}/C$ value, which may be obtained by scaling from the 10-km values, and the average photon number, $\mu$.

| | |
|---|---|
| 651-bit daylight sifted key with Bob's errors in red | 0000011001100010010000010110100**00**1100111001000111110000100011110010**11**<br>001001100000010010011001100010101101101011011111100001000011111000**11**1111101<br>111110**1**1110101111000010100100010010101110000110110000101100000100101 1<br>100101001110010100111110110110010000111100010000110**11**1011100011100001 01<br>001000100010001100101100111011101111111000001111101101100110**000**011100010 1<br>0100110**10**01101001101010001**0**101100000101000111011011110011011011110<br>100111100100100101010101100**0**110010101010100000110111100100100010<br>011111011100100111010001111001011100010001**0101**1101100010111100<br>0001111111010101011101000001111001100**1**011011101001101**0**111100011110101110<br>001000011000010101110010110110010110110 |
| Final 264-bit error-free, secret daylight key | 1100000011000001010001010010001011011111000000101001111001000011111<br>1011111111111011110011111011110010111111110000100101101000100000001111<br>1111011100100100101011110110110111001010010010101011110001010100111010111<br>0100000111101010101101010000011110000100111110011010001110111011001100<br>01001101010101110000011011100011010001000110000100110011000 |

Table 1: Sifted and final secret keys from a 1-s daylight quantum key transmission at 18:40:26 MDT on 4 October, 2001, which had an average photon number, $\mu = 0.29$, and an atmospheric quantum channel efficiency, $\eta_{opt} = 2.4\%$, resulting in 1,349 raw key bits, from which 651 bits were sifted, composed of 331 bits from the rectilinear basis and 320 bits from the diagonal basis. (The average photon numbers of each BB84 polarization state transmitted were, $\mu = 0.291$, 0.288, 0.291, and 0.288 for the "H", "V", "+45º", and "-45º" polarizations respectively.) Bob's sifted key contains 21 errors (shown in red), corresponding to $C \approx 5$, with 2 "H" errors ("H" transmitted but "V" received), 7 "V" errors, 7 "+45º" errors and 5 "-45º" errors, giving a sifted BER of $\varepsilon = 3.2\%$, which translates to a background radiance $\sim 0.2$ mW cm$^{-2}$ $\mu$m$^{-1}$ str$^{-1}$. Alice and Bob estimate that Eve's collision entropy on the sifted key is reduced from maximal by 40 bits to compensate for potential intercept-resend eavesdropping in the Breidbart basis on single-photon events, and by 170 bits to compensate for potential eavesdropping on multi-photon events, to 440 bits. This is further reduced by 155 bits to compensate for side information revealed to correct Bob's errors, by 2 bits of side information corresponding to a slight 47/53-bias towards 0s in the sifted key, and by a 20-bit safety factor to give a 264-bit final, error-free secret key on which Eve's expected information is $< 10^{-6}$ bits. During this transmission there was one multi-detector event, consistent with a data-background coincidence, which was discarded. We attribute the tendency for errors to cluster to atmospheric scintillation.